\documentclass{article}
\usepackage{PRIMEarxiv}

\usepackage{authblk,graphicx,amsmath,url,amsfonts,bm,subcaption,booktabs,nccmath,siunitx,ragged2e,multirow,caption,comment,footnote,xcolor}
\usepackage{hyperref,cleveref}
\hypersetup{colorlinks=true,linkcolor=blue,urlcolor=blue,citecolor=blue}

\title{A multi-patient analysis of the center of rotation trajectories using finite element models of the human mandible}

\author[a,b,*]{Torkan Gholamalizadeh}
\author[a]{Sune Darkner}
\author[b]{Peter Lampel Søndergaard}
\author[a]{Kenny Erleben}
\affil[a]{\textit{Department of Computer Science, University of Copenhagen, Copenhagen, Denmark}}
\affil[b]{\textit{3Shape A/S, Copenhagen, Denmark}}
\affil[*]{\textit{torkan@di.ku.dk}}

\begin{document}
\maketitle

\begin{abstract}
Studying different types of tooth movements can help us to better understand the force systems used for tooth position correction in orthodontic treatments. This study considers a more realistic force system in tooth movement modeling across different patients and investigates the effect of the couple force direction on the position of the center of rotation (CRot). The finite-element (FE) models of human mandibles from three patients are used to investigate the position of the CRots for different patients’ teeth in 3D space. The CRot is considered a single point in a 3D coordinate system and is obtained by choosing the closest point on the axis of rotation to the center of resistance (CRes). A force system, consisting of a constant load and a couple (pair of forces), is applied to each tooth, and the corresponding CRot trajectories are examined across different patients. To perform a consistent inter-patient analysis, different patients’ teeth are registered to the corresponding reference teeth using an affine transformation. The selected directions and applied points of force on the reference teeth are then transformed into the registered teeth domains. The effect of the direction of the couple on the location of the CRot is also studied by rotating the couples about the three principal axes of a patient’s premolar. Our results indicate that similar patterns can be obtained for the CRot positions of different patients and teeth if the same load conditions are used. Moreover, equally rotating the direction of the couple about the three principal axes results in different patterns for the CRot positions, especially in labiolingual direction. The CRot trajectories follow similar patterns in the corresponding teeth, but any changes in the direction of the force and couple cause misalignment of the CRot trajectories, seen as rotations about the long axis of the tooth.
\end{abstract}


\section{Introduction}

Finite element (FE) modeling is a widely used computational method for the analysis of the reactions to real-world physical effects such as forces and biomechanical responses occurring during treatments in medicine and dentistry \cite{savignano2019biomechanical} that attempts to solve partial differential equations numerically, based on reconstructing the desired geometry and discretizing the domain into a finite mesh, with minimum need for clinical trials of patients \cite{cervino2020application}. One of the main goals in FE-based modeling of the tooth and its supporting complex, i.e., periodontal ligament and adjacent alveolar bone, is to improve tooth movement prediction performance in orthodontic treatments. Compared to clinical studies that aim, among other things, to assess the accuracy of digital planning in dentistry \cite{lavorgna2019reliability} or investigate anatomical characteristics for optimal occlusions \cite{sambataro2018upper} using different patient scans, computational modeling allows for more efficient plannings in orthodontic treatment for correcting dentofacial anomalies using more generic solutions.

Tooth movement modeling in an uncontrolled tipping scenario using a perpendicular loading system has been widely used for initial tooth movement simulations due to its simplicity \cite{gholamalizadeh2020mandibular}. However, in addition to the uncontrolled tipping movements, different tooth movements such as the pure translation, crown/root tipping, intrusion/extrusion, or a combination of them are typically required for tooth position correction in orthodontic treatments. Each of these movements can be described based on the position of the center of resistance (CRes) and center of rotation (CRot) with respect to the tooth geometry \cite{nyashin2016centre,meyer2010does}. For example, a pure translation, root tipping, and crown tipping can result in CRots located at infinity, crown, and root apex, respectively. The CRes of a tooth within its supporting complex can be seen as the center of the mass of a free rigid body \cite{viecilli2013axes}. To be more specific, the tooth CRes is a point at which applying any forces would always result in a pure translation of the tooth \cite{nyashin2016centre}. Although the location of the CRes has been investigated in the literature either experimentally in vivo \cite{sia2007determining,sia2009experimental,yoshida2001experimental}, analytically \cite{van2013analytical,provatidis2001analytical}, or computationally using finite-element (FE) models \cite{meyer2010does,viecilli2013axes,oh20193}, estimating the exact position of the CRes is challenging as it depends on the root shape, the anatomy and mechanical properties of the supporting complex \cite{schneider2002numerical,meyer2010does}, the direction of the tooth movement and force \cite{schmidt2016comparison,meyer2010does,melsen2007importance,nyashin2016centre,liao2016biomechanical}. More specifically, since the CRes may not exist as a single point in 3D space \cite{viecilli2013axes} due to the asymmetric geometries, it is proposed to use a volume of CRes instead \cite{oh20193}.

The CRot is another important concept in tooth movement analysis and is a point at which the movement of the tooth can be defined as pure rotation. Different approaches can be used for determining the position of the CRes and CRot. One simple approach for determining the tooth CRes is exerting a couple in different directions. A couple consists of a pair of forces with the same magnitude and parallel line of action, but in opposite directions where they are not collinear \cite{smith1984mechanics}. Therefore, utilizing only a couple results in a pure rotation in the tooth where the position of the CRot coincides with the position of the CRes.

The CRot position depends on the load system applied to the tooth. The loading system can result in some moments and forces in different directions, and due to its complexity in a 3D coordinate system, it can be decomposed into three planes of the tooth. Hence, tooth movement can be analyzed by decomposing each moment-to-force ratio ($M:F$) on each principal plane of a tooth, where each $M:F$ on a plane can be defined based on the force vectors located on the plane and moments perpendicular to the studied plane \cite{savignano2016nonlinear}. 

A classical theory on the relation between the applied load and the type of the tooth movement is \textit{Burstone's formula} \cite{burstone1962biomechanics}. In this theory, the effect of the various moment-to-force ratios ($M:F$) is studied on the position of the CRot for a canine with a parabolic root, while the force is applied perpendicular to the long axis of the tooth at the bracket level. The study introduces the following formula
\begin{equation} 
\label{eq:burstone}
\nonumber M:F = \frac{0.068 \times h^2}{D},
\end{equation}
\noindent where $h$ is the distance from the alveolar crest to apex, and $D$ is the perpendicular distance between the CRes and CRot. Note that $M:F$ has a unit of millimeters where $F$ corresponds to the magnitude of the applied force and $M$ denotes the moment of the couple $m_c$ exerted to the tooth in order to counteract the moment caused by the applied force $m_f$ \cite{smith1984mechanics}. According to this theory, specific values of $M:F$ ratios always correspond to specific types of tooth movement. That is, the location of the CRot for a specific tooth is only dependent on the moment-to-force ratio of that tooth.

The classical theory has been widely studied by using FE models \cite{melsen2007importance,cattaneo2008moment,savignano2016nonlinear}. Cattaneo et al.  \cite{cattaneo2008moment} study the influence of $M:F$ on the same force magnitude and analyze the effect of applying various load magnitudes on a mandibular premolar and canine teeth with a constant $M:F$. The study reports that a constant $M:F$ with different force magnitudes can result in different types of tooth movements. However, this finding does not follow Burstone's formula, and this is due to considering a nonlinear material model for the PDL layer where the location of the CRot is dependent on both the $M:F$ and the magnitude of the applied load to the teeth.

Different studies have utilized different $M:F$ ratios for the pure translation of various teeth. For example, unlike the generally accepted $M:F$ of 10 and 12 for pure translation of the premolar and canine, Cattaneo et al. \cite{cattaneo2008moment} suggest $M:F$ of 9 and 11, respectively, for a mandibular premolar and a canine. Also, it is recommended \cite{liao2016biomechanical} to avoid using a universal $M:F$ and CRes due to the patient-specific geometries, reporting $M:F$ of approximately 8.8, 9.7, and 10.2, to simulate the pure translation of maxillary first premolar, lateral incisor, and canine. 

Unlike the abovementioned studies that are limited to single-plane analysis, Savignano et al. \cite{savignano2016nonlinear} investigate the effect of the force system directions on the tooth movement of a maxillary first premolar by performing analysis on principal planes of the tooth. More specifically, the mesiodistal, buccolingual, and occlusal planes are used to study the location of the CRot using the projected axis of rotation on these planes for various $M:F$ and directions. This study suggests a nonlinear relation between the direction of the force system and the projected CRots. However, the mentioned study is limited to a single tooth analysis with a limited load magnitude range applied to the CRes due to using a linear elastic material model for the PDL layer. Although two patients are considered in a recent work \cite{savignano2020three}, an inter-patient analysis is missing in general.

The existing studies do not represent the effect of geometrical variations on the $M:F$ ratios and the position of the CRot under the same analysis setup, boundary conditions, and loading system for different patients. To this end, in this work, we consider patient-specific full dentition computational models of three human mandibles to investigate the position of the CRots in different patients’ teeth. For a comprehensive and consistent inter-patient analysis, we register different patients’ teeth to a corresponding reference tooth of the first patient using an affine transformation and use this transformation to map the selected force directions and force application points on the reference teeth into the registered teeth domain. Moreover, we model the clinical forces with no specific limitation in the force magnitudes by using a hyperelastic material model for PDL tissue and investigate the position of the CRot in a 3D space to better represent the resulting rotation. Finally, we assess the influence of the couple directions on the positions of the CRots of a specific tooth and show that equally rotating the direction of the couple about the three principal axes results in different patterns for the CRot positions, especially in labiolingual direction. Modeling different types of tooth movements using a more realistic force system, i.e., tipping and couple forces applied to the tooth crown, and investigating the effect of these couple forces on the CRots can help us to better estimate the force systems needed to achieve the desired tooth movement.

\section{Materials and methods}

This section first reviews the details of the FE models used in this study including the geometry reconstruction, mesh generation, boundary conditions, contact definitions, and material models. Next, it focuses on describing an approach for setting up a consistent loading condition for different patients' teeth, both for intra- and inter-patient analysis. Finally, it briefly specifies how the CRot is computed and illustrated in 3D.

\subsection{Geometry reconstruction}

Three patient-specific and anatomically accurate FE models of the human mandible composed of mandibular teeth, corresponding PDLs, and bone are considered in this work. To have enough geometrical variations in the dataset, the three patients' scans are chosen of various crown heights, root lengths, and teeth sizes. The scans are used from the 3Shape A/S private dataset collected from different clinics by orthodontists as a part of treatment. More specifically, the anonymized Cone-Beam Computed Tomography (CBCT) scans of patients stored in Digital Imaging and Communications in Medicine (DICOM) format used in this study contain no sensitive personal information including gender and age or details of the scanner device utilized for image acquisition. 

The utilized patient scans labeled as \textit{Patient 1}, \textit{Patient 2}, and \textit{Patient 3} have an isotropic voxel size of \SIlist{0.3}{\milli\metre}, \SIlist{0.3}{\milli\metre}, and \SIlist{0.15}{\milli\metre}, respectively. For a consistent and detailed geometry reconstruction, a cropped region of interest (ROI) of every patient scan including full dentition and mandible is upsampled to an isotropic voxel size of \SIlist{0.15}{\milli\metre}. Further details of the utilized scans can be found in Table \ref{table:scan_info}. Note that the patient-specific geometries are reconstructed by importing DICOM files to 3DSlicer \cite{fedorov20123d}, resampling the ROIs, and segmenting the teeth and bone geometries in CBCT scans using a semi-automatic watershed algorithm.

\begin{table}[!t]
\renewcommand{\arraystretch}{1.25}
\caption{Detail of the utilized scans.}
\resizebox{\textwidth}{!}{%
\begin{tabular}{lccccc}
\hline
\multirow{2}{*}{\begin{tabular}[c]{@{}l@{}}Patient\\  ~~ID\end{tabular}} & \multicolumn{3}{c}{Dimensions} & \multicolumn{2}{l}{Slice thickness (mm)} \\ \cline{2-6} 
 & Scan size & ROI size & Resampled ROI size & Original & Resampled ROI \\ \hline
Patient 1 & $400\times400\times280$ & $338\times265\times140$ & $776\times530\times280$ & 0.3 & 0.15 \\
Patient 2 & $400\times400\times280$ & $335\times220\times172$ & $670\times440\times344$ & 0.3 & 0.15 \\
Patient 3 & $532\times532\times540$ & $534\times435\times338$ & $534\times435\times338$ & 0.15 & 0.15 \\ \hline
\end{tabular}%
}
\label{table:scan_info}
\end{table}

The accuracy of the annotated scans is verified by a clinical expert and any required modifications are applied to the segmented regions accordingly. A general criterion for the verification is precise segmentation of roots, crowns, and bone cervical regions including the teeth sockets, and the miss-annotated regions indicated by the expert are revised until reaching the criterion. The segmented geometries are then exported as surface meshes in STL file format. Since the resolutions of the original scans are not high enough to properly represent the thin layer of the PDL, annotating the PDL layer using the available CBCT scans is not feasible. The width of the PDL layer is a tooth root dependent factor which can vary between \SIlist{0.15}{\milli\metre} and \SIlist{0.38}{\milli\metre} with an average suggested as \SIlist{0.2}{\milli\metre} \cite{li2018orthodontic}. This average thickness is used in the literature \cite{li2018orthodontic,barone2017mechanical,seo2021comparative} to generate the PDL geometries by uniformly extruding the teeth roots in Meshmixer \cite{schmidt2010meshmixer}. Therefore, we first re-mesh the teeth and bone geometries using a uniform and adaptive mesh, respectively. Next, the obtained bone surface is offset by \SIlist{0.2}{\milli\metre} in the reverse direction of the surface normals to create the required space for the PDL layers in between the teeth roots and the regenerated bone. Finally, the PDL geometries are produced by extruding the teeth roots, filling the generated gap in between the teeth and bone. It should be noted that the generated surface mesh for each PDL geometry includes a uniform mesh with two elements in the PDL thickness. The reconstructed geometries have been made publicly available and can be obtained from Electronic Research Data Archive at the University of Copenhagen under the OpenJaw Dataset (\url{https://doi.org/10.17894/ucph.04e91c97-5c5d-45d7-afd3-c5e0b5953f58}).

After preparing the surface meshes, high-quality tetrahedral meshes are generated by preserving the surface meshes using TetGen \cite{si2015tetgen}. The mesh quality required for an FE analysis can vary depending on the application and utilized numerical methods \cite{shewchuk2002good}. In general, a regular tetrahedron has the highest mesh quality, and the main guideline is to avoid using low-quality tetrahedra with small or large dihedral angles, as they can affect the accuracy of the numerical methods \cite{shewchuk2002good}. Therefore, we use an upper limit constraint of 1.2 for the radius-edge ratio, as the ratio of the circumscribed sphere's radius and the shortest edge of the linear tetrahedral meshes. Additionally, the quality of the generated volumetric meshes are assessed using four different quality measurements, i.e., the volume-edge ratio \cite{liu1994relationship,misztal2013multiphase}, radius-edge ratio \cite{baker1989element}, radius ratio \cite{freitag1997tetrahedral,caendish1985apporach}, and volume-area ratio \cite{freitag1997tetrahedral}. The volumetric meshes are then imported into the FEBio software package \cite{maas2012febio}, which is open-source software for nonlinear FE analysis in biomechanics, to set up a reproducible FE model for each patient. A summary of the utilized mesh properties is provided in Table \ref{table:model_params}.

\begin{table}[!t]
\centering
\renewcommand{\arraystretch}{1.35}
\caption{Summary of the materials and mesh properties.}
\label{table:model_params}
\resizebox{\textwidth}{!}{%
\begin{tabular}{l|l|cc|lclll}
\hline
\multirow{2}{*}{Domain} & \multirow{2}{*}{Material model} & \multicolumn{2}{c|}{Material properties} &  &~~ Mesh properties &  & \multicolumn{1}{l}{} & \textbf{} \\
 \cline{3-4} \cline{5-9} 
 &  & \begin{tabular}[c]{@{}c@{}}Young's modulus\\ 
 (MPa)\end{tabular} & \begin{tabular}[c]{@{}c@{}}Poisson's ratio\\ ($ - $)\end{tabular} &  & \begin{tabular}[c]{@{}c@{}}Surface mesh \\ Edge length (mm)\end{tabular} &  & \multicolumn{1}{l}{\begin{tabular}[c]{@{}l@{}}Volumetric mesh\\  Number of elements $\dagger$ \end{tabular}} & \textbf{} \\ \hline
Tooth  & Rigid body & $ - $ & $ - $ &  & 0.4  &  & 14$\pm$5 K & \multirow{4}{*}{} \\
Bone & Isotropic elastic & $1.5\times10^{3}$ & 0.3 &  & \begin{tabular}[c]{@{}c@{}}Adaptive mesh\\ (from 0.4 to 2)\end{tabular} &  & 3,261$\pm$329 K &  \\ \cline{3-4}
 &  & $C_1$ (MPa) & $C_2$ (MPa) &  &  & \multicolumn{1}{l}{} &  &  \\ \cline{3-4}
PDL & Mooney-Rivlin $^*$ & 0.011875 & 0 &  & 0.1 &  & 127$\pm$41 K &  \\ \hline
\end{tabular}%
}
\begin{flushleft}
\footnotesize
$^{*}$ $C_2 = 0$ reduces the Mooney-Rivlin material to uncoupled Neo-hookean. The values assigned for $C_1$ and $C_2$ correspond to the Young’s modulus and Poisson's ratio of \SIlist{0.0689}{\mega\pascal} and $0.45$, respectively. \\
$\dagger$ The values are shown as the mean $\pm$ standard deviation of the number of elements across all cases.
\end{flushleft}
\end{table}

A mesh convergence study is performed to obtain an optimal mesh size for the FE models \cite{gholamalizadeh2020mandibular} by iteratively increasing the number of elements per step with a factor of 2. We continue the process until the relative stress error does not exceed 4\% of maximum von Mises stress. This study is concerned with smooth bone meshes where there are no sharp elements on the meshes. Therefore, the model does not experience extreme local stresses in a single node or element. However, in nonsmooth geometries with sharp edges, more robust methods can be applied to avoid any extreme local stresses as outliers \cite{zmudzki2008stress,roda2021computerized}. Also, note that using more elements in the PDL thickness would exponentially increase the total mesh size of the full jaw models and the required computational time. Besides, based on our experience, using smaller elements would result in extreme element distortion, causing problems in FE model convergence especially for higher load values. Similar behavior is also seen in \cite{ortun2018approach} for hexahedral elements.

\subsection{Material properties}

We assume teeth as rigid bodies with six degrees of freedom to simplify the proposed model. Also, we assume the center of mass as the CRes for each tooth which can automatically be calculated using FEBio user-defined options \cite{FEBioUserManual}. Since the deformation of the bone tissue is negligible under the orthodontic forces used in this study, no distinction is made for the cortical and trabecular bone \cite{ziegler2005numerical,qian2009deformation}, and isotropic elastic material model is used for the homogeneous bone geometry. Due to the importance of the PDL tissue in transferring loads from the tooth to the alveolar bone \cite{ortun2018approach,cattaneo2005finite}, a Mooney-Rivlin Hyperelastic (MRH) material model is used to simulate its nonlinear behavior \cite{qian2009deformation,uhlir2016biomechanical}. Furthermore, for the simplicity of the model and following the literature \cite{ortun2018approach,oh20193,savignano2016nonlinear,savignano2020three}, the gingival tissue is discarded from our computational model due to its extremely low elasticity modulus compared to that of the other tissues, i.e., PDL, bone, and tooth.

\subsection{Boundary condition and contact definition}

A \textit{rigid contact} is defined between rigid teeth and PDLs, and a tied \textit{facet-on-facet} contact with an augmented Lagrangian method is used in PDL-bone interfaces to model adhesion in these two interfaces, as their corresponding surfaces do not have any sliding or separation. Other parameters, such as boundary conditions, edge length of the elements, and material properties are chosen as mentioned in Table \ref{table:model_params}. Also, a Dirichlet boundary condition is applied to all nodes at the bottom surface of the bone in all directions to fix the jaw model.

\subsection{Loading conditions for different scenarios}

\begin{figure}[!h]
\centering
\includegraphics[scale=1.55]{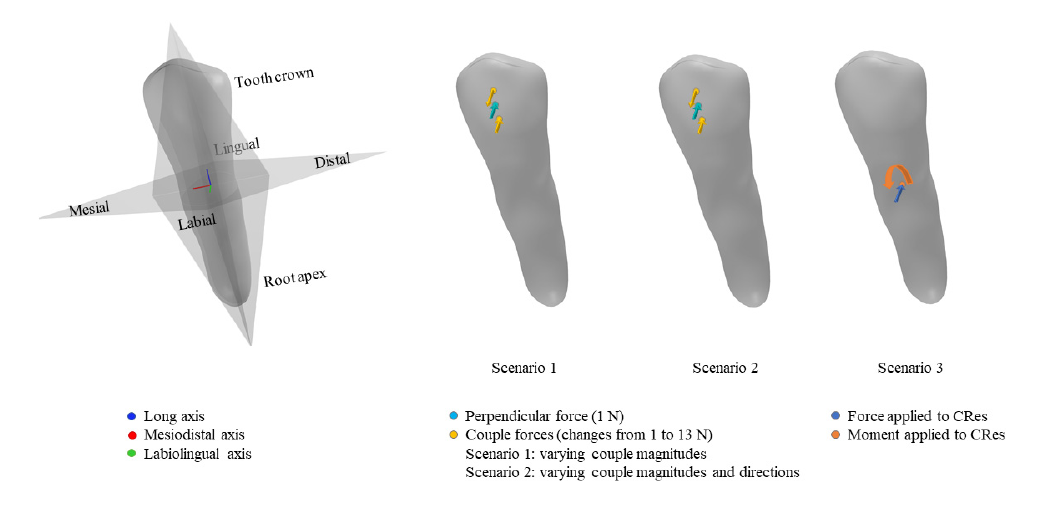}
\caption{{\bf The studied scenarios.} An illustration of the principal axes and planes of premolar and loading conditions used in three different scenarios.}
\label{fig:scenarios}
\end{figure}

Three different scenarios with varying loading conditions are studied in this work, as shown in Fig \ref{fig:scenarios}. In the first scenario, various $M:F$ values are examined for different patients' teeth using a consistent loading system. In the second scenario, the influence of the direction of the couple is studied on a specific tooth. As our third scenario, the loading conditions are applied to the CRes, which makes us able to compare our simulation results with those obtained by Savignano et al. \cite{savignano2016nonlinear,savignano2020three}.

\subsubsection{Scenario 1: Consistent loading conditions for intra- and inter-patient analysis} \label{scenario1}

In the first scenario, to perform a systematic intra- and inter-patient analysis and to avoid potential errors caused by the selection of the load points and force directions, different teeth of each patient are registered to a reference tooth of the same type. To do so, the left teeth of one of the patients (here, \textit{Patient 1}), with the \textit{Universal Numbering (UNN)} ID of 18 through 24, are selected as the reference teeth. Next, the mandibular teeth of all other patients on both sides, as well as the right mandibular teeth of \textit{Patient 1}, are all registered to the corresponding reference teeth. This allows us to analyze more tooth models of the same type and study the position of the CRots for different teeth of different patients.

Before any tooth registration, the jaw models of different patients are aligned based on rigid body transformation using the Iterative Closest Point (ICP) algorithm \cite{chen1992object,besl1992method}. Moreover, the Coherent Point Drift (CPD) algorithm \cite{myronenko2010point} is used for teeth alignment based on an affine transformation of the corresponding left/right teeth of different patients. The registration processes are all performed in MATLAB by considering vertices of each surface mesh as a set of a point cloud. Compared to the ICP algorithm, CPD is computationally intensive, yet it is more robust to noise and outliers. Therefore, we prefer using CPD for teeth registration which includes fewer data points than jaw models. It should also be noted that all teeth on the right side of the mandibles, with the UNN of 25 through 31, are horizontally flipped with respect to the sagittal plane of each patient's jaw before applying the affine transformation using the CPD method. This is to ensure that the labial surfaces of the registered teeth are properly aligned to the labial surfaces of the reference teeth.

After registering all patients’ teeth to the reference teeth, three points, at which the force and couple are applied, are selected on each reference tooth. One point is set for tipping force, and two points are used for the couple. The center of the middle third of the crown is set for the tipping load, and two points are selected with a \SIlist{0.5}{\milli\metre} distance below and top of the tipping load point for the labiolingual and linguolabial couple, respectively. The normal vector of the reference mesh at the tipping point is used for both directions of the force and the couple. More specifically, the direction of the computed normal vector is used for the linguolabial couple force on top, and the opposite direction is used for both the tipping force and the labiolingual couple. \textit{Nodal loads} are utilized to simulate the mentioned forces in the FEBio framework.

\begin{figure}[!h]
\centering
\includegraphics[scale=1.55]{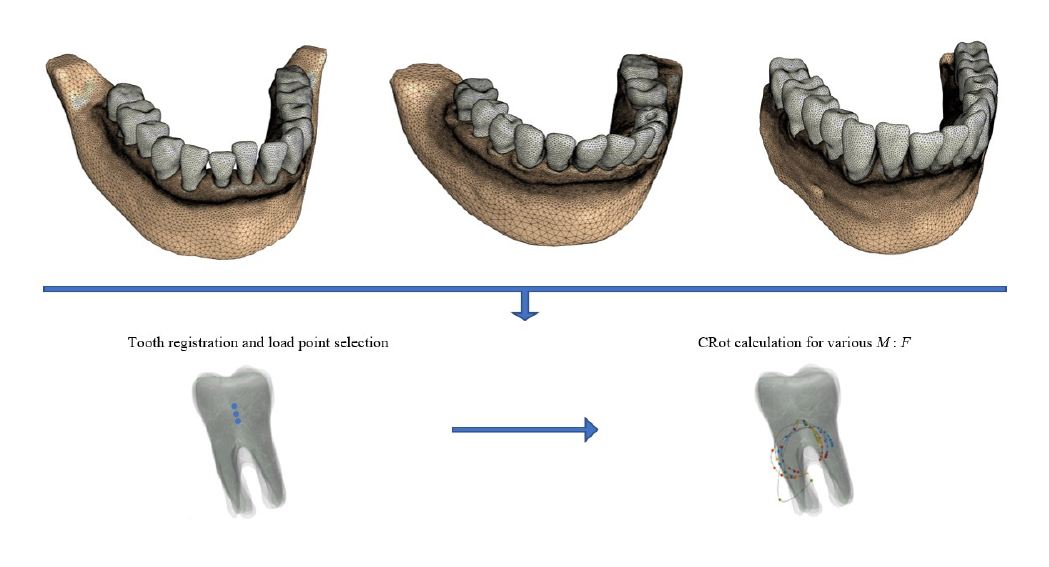}
\caption{{\bf Tooth registration and CRot calculations pipeline.} An illustration of the utilized method for calculating the CRots using finite-element models of mandibles from three patients under consistent loading conditions.}
\label{fig:pipeline}
\end{figure}

The corresponding three load points are then obtained based on finding the closest mesh points in each registered tooth mesh. These points, as well as the direction of the forces, are finally transformed back to the original coordinates of each patient’s tooth. This would ensure unbiased loading conditions for the intra- and inter-patient analysis. Fig \ref{fig:pipeline} displays how registration is employed for defining consistent loading conditions and calculating the CRots using FE models of mandibles from three patients.

Multi-patient analyses are conducted on different teeth to evaluate the effect of the couple’s magnitude on counteracting the moment produced by the tipping load. In this scenario, the simulations are performed with a constant tipping force of \SIlist{1}{\newton} and varying couple magnitudes of \SIlist{1}{\newton} to \SIlist{13}{\newton} with \SIlist{1}{\newton} intervals. Note that the applied force points are fixed for both the couple and the tipping load. Hence, increasing the load magnitude of the couple will increase the resulting moment on the tooth according to the force-moment relation $M = F \times d$, where $d$ is the distance between the applied force point and the CRes of the tooth. Finally, the corresponding CRots are computed for each patient’s teeth and the same transformations are used to align different teeth of the same type and their CRots.

\subsubsection{Scenario 2: The effect of the couple's direction on the position of the center of rotation} \label{scenario2}

In the second scenario, as an example, the left premolar of \textit{Patient 1} is selected to study the influence of the couple's direction on the position of the CRot. The three load points, the initial load directions, and the direction and magnitude of the tipping load remain the same as in the first scenario. The couple’s directions are rotated about the mesiodistal, the labiolingual, and the long axis of the tooth with different degrees varying from 0 to 90 degrees with 10-degree increments. To investigate the trajectories of the CRots in 3D space, for each direction, we conduct experiments with incremental couple magnitudes of \SIlist{1}{\newton} to \SIlist{13}{\newton} with intervals of \SIlist{0.0625}{\newton}. This generates a dense representation of the CRot trajectories in 3D space.

\subsubsection{Scenario 3: The effect of the force's direction on tooth movement} \label{scenario3}

As our third scenario, we investigate the influence of the force’s direction on tooth movement and the position of the CRot, and compare our results with those obtained by Savignano et al. \cite{savignano2016nonlinear,savignano2020three}. More specifically, the effect of the force direction is investigated at the CRes of the left premolar of \textit{Patient 1}, while the load direction changes in one of the three principal planes of the tooth, and a perpendicular moment is applied to the studied plane. Additionally, instead of using the couple for generating the moments on the tooth crown, we follow the same scenarios of the aforementioned study and use a load and a moment applied to the CRes, using \textit{prescribed rigid force constraint} and \textit{prescribed rigid torque constraint}, respectively, in FEBio framework. The direction of the load changes with 10-degree increments, and the $M:F$ changes from -12 to 12 with \SIlist{2}{\milli\metre} intervals.

\subsection{Computing the center of rotation}

The axis of rotation is computed for each tooth by considering displacement vectors of two nodes arbitrarily selected on the tooth crown. To this end, the intersection line of the perpendicular bisector planes of the displacement vectors is obtained. Any point on the intersection axis can be assumed as the CRot in 3D space, and to find a unique CRot in 3D, the point with the closest distance to the CRes of the tooth is selected on the rotation axis. This allows for representing the axis of the rotation in 3D as a single point and for better analyzing the influence of the couple directions on the location of the CRot with respect to the CRes.

\section{Results}

\begin{figure}[!h]
\centering
\includegraphics[scale=1.45]{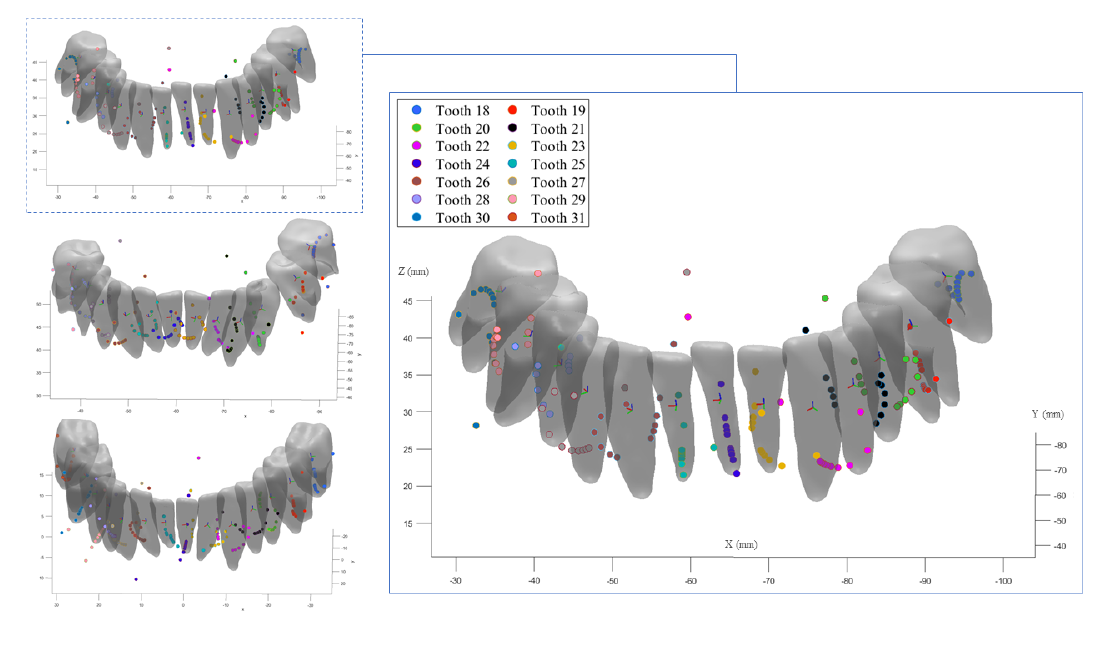}
\caption{{\bf Calculated CRots.} Calculated CRots for different couple magnitudes changing from \SIlist{1}{\newton} to \SIlist{13}{\newton} with \SIlist{1}{\newton} intervals shown in 3D space for all teeth with transparent teeth geometries, \textbf{Left:} Results of the three different patients. \textbf{Right:} A closeup view of the results for \textit{Patient 1}.}
\label{fig:crot_all}
\end{figure}

We conduct our experiments on a 3.4 GHz processor with 64 GB of RAM, which takes about an hour to solve the FE model. First, we focus on the analysis of the position of the CRots for various couple magnitudes of all patients’ teeth. Second, the influence of the couple directions is studied on the position of the computed CRot of the left premolar of \textit{Patient 1}. Third, we assess the influence of the force direction with fixed moment direction on the position of the CRots. In all the experiments, the position of the closest point to the CRes on the rotation axis is considered as a single point CRot in 3D coordinate system for each tooth. Fig \ref{fig:crot_all} illustrates the position of the CRots of each tooth for varying magnitudes of the couple from \SIlist{1}{\newton} to \SIlist{13}{\newton} with \SIlist{1}{\newton} increments. As can be seen, the CRots follow parabolic shapes with varying slopes from one tooth type to another.

\begin{figure}[!h]
\centering
\includegraphics[scale=1.15]{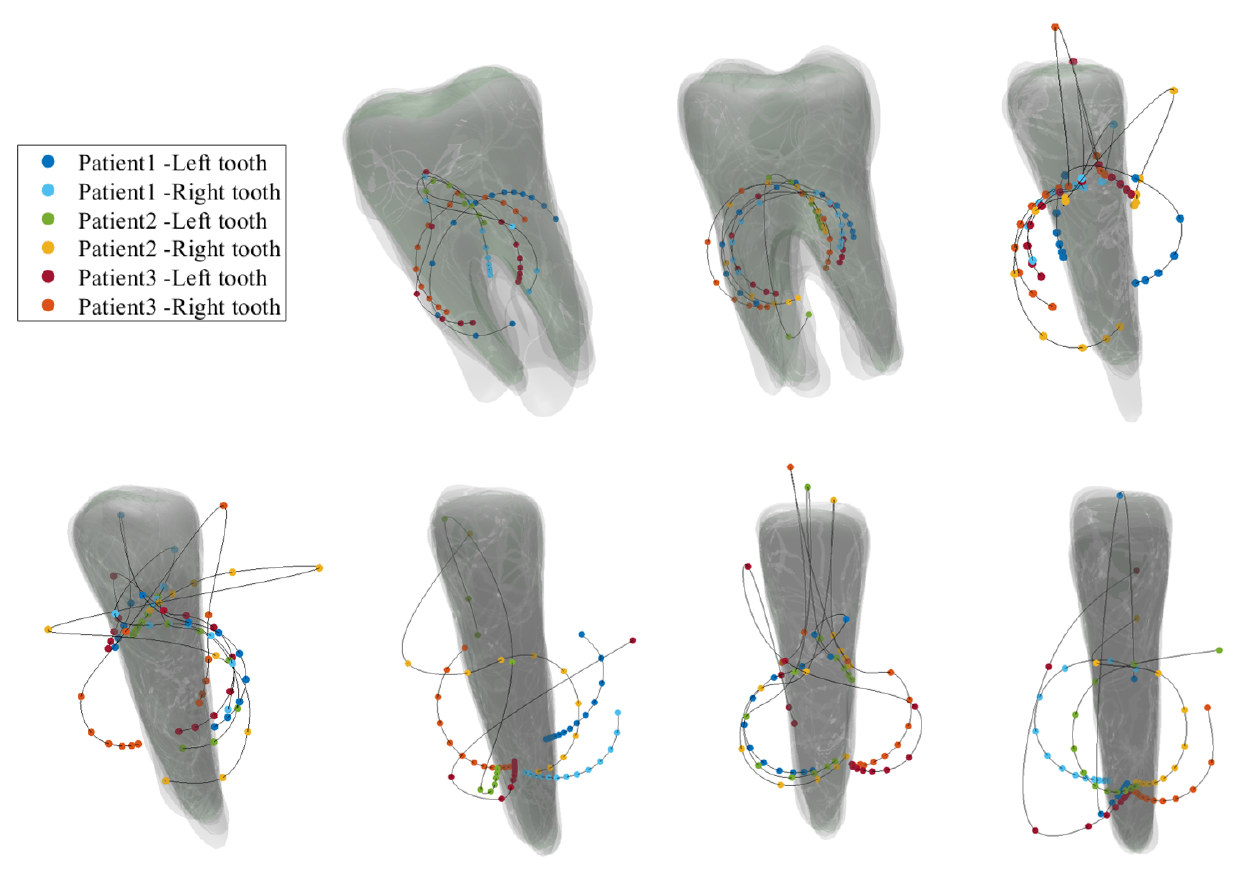}
\caption{{\bf CRots trajectories of registered teeth in scenario 1.} Trajectories of the obtained CRots for the different patients’ teeth registered to the reference teeth of the same types and shown as transparent teeth. \textbf{Top-row:} The second molars, the first molars, the second premolars. \textbf{Bottom-row:} The first premolars, the canines, the lateral incisors, and the central incisors. A consistent loading condition is applied to all teeth of the same types, and the magnitude of the couple changes from 1N to 13N. Note that a few extreme outliers are discarded for a better illustration purpose.}
\label{fig:result_reg}
\end{figure}

To better represent the inter-patient results, the utilized patients’ teeth and their CRots were registered to the corresponding reference teeth. Fig \ref{fig:result_reg} shows the registered teeth geometries, with transparent shapes located on either left or right side of the mandible, and their CRots for the utilized patients. Note that the third molars were missing in all cases. Hence, the different teeth types are represented from the second molars to the central incisors. As can be noted, the CRot trajectories experience similar patterns in each tooth type.

\begin{figure}[!h]
\centering
\includegraphics[scale=1.4]{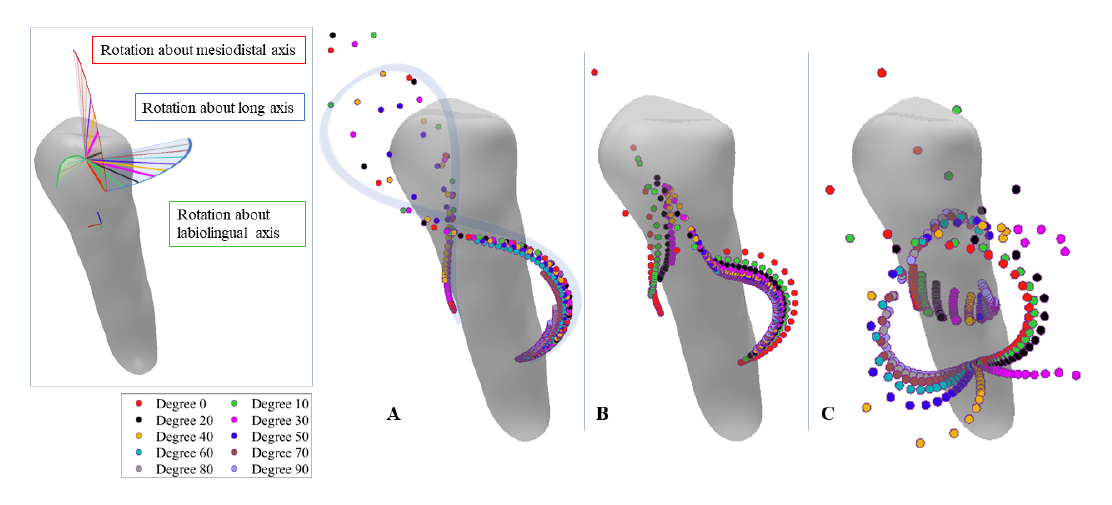}
\caption{{\bf Effects of the couple's directions on the CRots trajectories in scenario 2.} Directions of the couple rotated about the tooth principal axes with various angles and the corresponding CRots for different couple magnitudes. \textbf{Left:} An illustration of how the couple forces are rotated about mesiodistal, long axis, and labiolingual of the tooth. \textbf{Right(A to C):} The corresponding CRots for the rotated couple about mesiodistal, long axis, and labiolingual of the left premolar of \textit{Patient 1}, respectively. Note that the initial direction of the couple and tipping load are set based on the surface normal direction.}
\label{fig:couple_rotated}
\end{figure}

In practice, more CRot data points are required for a better illustration of the trajectory curves in 3D space. This in turn requires sampling more data points by using smaller intervals of the couple magnitudes. As mentioned before, we conduct this experiment using the left premolar of \textit{Patient 1} and investigate the CRot trajectories influenced by the couple's directions based on smaller increments in the magnitude of the couple. More specifically, the directions of the couple are rotated about the mesiodistal, labiolingual, and long axes of the tooth from 0 to 90 degrees with 10-degree increments. Fig \ref{fig:couple_rotated} illustrates the CRot trajectories for different directions of the couple rotated about the principal axes of the tooth.

\begin{figure}[!h]
\centering
\includegraphics[scale=1.5]{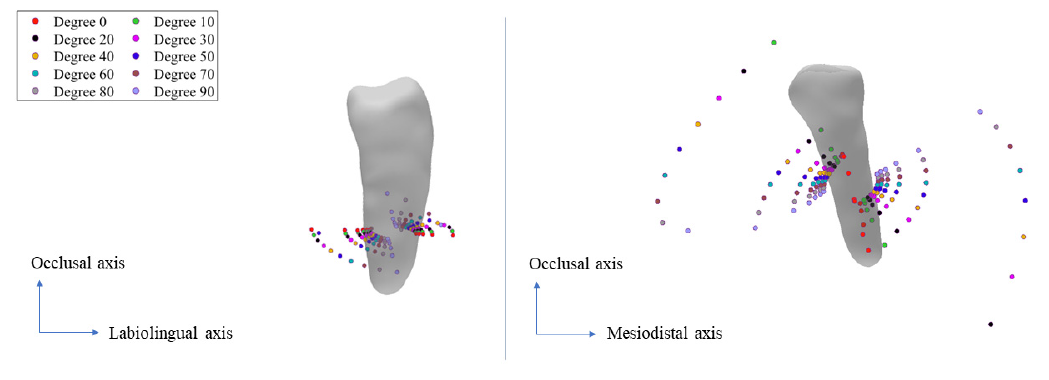}
\caption{{\bf The influence of the force direction on the position of the CRots in scenario 3.} The force and moments are applied directly to the CRes of the left premolar of \textit{Patient 1}. \textbf{Left:} The CRots of the forces rotating from long axis towards the labiolingual direction of the tooth. \textbf{Right:} The CRots of the forces rotating from the mesiodistal axis towards the long axis of the tooth. Note that each color corresponds to a different force direction, the $M:F$ changes from \SIlist{-12}{\newton} to \SIlist{12}{\newton}, and the distance between the CRots of the same color changes nonlinearly while the $M:F$ values increment linearly.}
\label{fig:savig}
\end{figure}

As our third analysis, we examine the effect of the force direction on the position of the CRot by considering three principal planes of the tooth. We use the same scenarios followed by Savignano et al. \cite{savignano2016nonlinear}, where the force and moment are applied to the CRes with a force of 1N magnitude. Fig \ref{fig:savig}, shows the effect of the load’s direction on the position of the CRot, where the rotation range of the direction of the force system can vary from one axis of the tooth to another with 10-degree increments. As it can be seen in both subfigures, the distance between the CRots of the same direction changes nonlinearly with linear increments of the $M:F$. The nonlinear variation increases when the force direction is rotated from the mesiodistal or labiolingual axis direction towards the long axis direction. Additionally, the distance of the CRots associated with the same $M:F$ ratios changes nonlinearly for linear increments of the rotation degree in 3D space. This pattern is also shown by Savignano et al. \cite{savignano2016nonlinear} on 2D planes.

\section{Discussion}

\subsection{Scenario 1}

As illustrated in Fig \ref{fig:result_reg}, the computed CRots for different patients follow similar trajectories for the teeth of the same types. It can also be seen that linear increments in the magnitude of the couples nonlinearly increases the distance between the neighboring CRots of each tooth. The same behavior was reported by Savignano et al. \cite{savignano2016nonlinear,savignano2020three}. Moreover, the CRot trajectories begin from the CRes of the tooth, and by increasing the $M:F$, the trajectories move towards the crowns of the teeth, and then, they change their directions to the back of the teeth. A schematic curve of the aforementioned path can be seen in part A of Fig \ref{fig:couple_rotated}.

The turning points can be found as outlying points on the top of each tooth with the highest distance from their neighboring CRots. These points are associated with specific $M:F$ values where the direction of the tooth movement changes in the opposite direction. In other words, in the turning point, the generated moment by the couple counteracts the generated moment caused by the tipping load. In addition, it can be observed that, in some cases, the CRot trajectories are not well aligned, i.e., they seem to be rotated about the long axis of the tooth. This could be due to the registration errors in matching the teeth models of different patients to each other, most likely caused by the size and high geometry variations of teeth.

\subsection{Scenario 2}

The investigation of the effects of the direction of the loading conditions on the CRot trajectories in Fig \ref{fig:couple_rotated} reveals that although rotating the couples about each axis results in similar CRot trajectories for different rotation angles, equally rotating the direction of the couple about the three principal axes results in different patterns for the CRot positions, especially in labiolingual direction. Besides, by increasing the magnitude of the couples in a specific direction, the distances between two neighboring CRots changes nonlinearly which is in line with the results of the state-of-the-art \cite{savignano2016nonlinear,savignano2020three}, indicating that changing the $M:F$ nonlinearly affects the position of the CRots.

As it can be observed in Fig \ref{fig:couple_rotated}, the CRot trajectories almost follow the same $\infty$-shaped curve by the rotation of the couple direction about the mesiodistal and long axes of the tooth with different angles (subfigure A and B). However, the sequence of the CRot points stops earlier when increasing the rotation angle. The situation in rotating the couple's direction about the labiolingual axis of the tooth looks a bit different. That is to say, the CRot trajectories follow different $\infty$-shaped curves which are seemed to be rotated about the long axis.

Considering the obtained results in Fig \ref{fig:result_reg} and subfigure C of Fig \ref{fig:couple_rotated}, one can deduce that any changes in the direction of force and couple can result in the misalignment of the CRot trajectories, and specifically, rotation of the CRots trajectories about the long axis of the tooth. In fact, the latter case shows how such systematic changes can affect the trajectory of the CRots, from which we can infer the reason for the misalignments that occurred in the former case. This behavior is somehow seen in a related work \cite{savignano2016nonlinear} where the CRot points scatter in the mesiodistal-long axes plane seems more incoherent than the other cases.

In this study, the direction of the load and couples are determined based on the normal of the tooth surface at the underlying region. As a result, the normal direction is not necessarily parallel to the labiolingual direction of the tooth. Therefore, in contrast to the literature where the force system is simplified in 2D planes by assuming force and couple directions parallel to the labiolingual direction, we investigate a more realistic clinical force system and study the influence of the couple's direction in the tooth movement.

\subsection{Scenario 3}

This work benefits from using more realistic material models and scenarios, such as applying the forces on the tooth crown, for better modeling of the forces in clinical environments. A hyperelastic material model is used here to model the nonlinear behavior of the PDL tissue under orthodontic forces which allows us to model different scenarios without any prior assumptions for the range of the applied load. Furthermore, we propose using 3D space for representing the position of the CRot for the exerted force and moment to help with better analysis and understanding of the results in real-world applications. In contrast, Savignano et al. \cite{savignano2016nonlinear,savignano2020three} have represented the results in 2D planes and applied the loads and moments on the CRes, which is difficult to precisely be determined for different teeth. Besides, they have utilized a linear material model to the PDL tissue which confines the range of the applied load \cite{kawarizadeh2004correlation}.

\subsection{Future directions}

Although an accurate geometry of the PDL can be obtained in vitro using micro-computed tomographies, these methods are known to be highly invasive and can only be applied to dead specimens. In contrast, reconstructing the PDL layer using CBCT scans obtained in vivo is a challenging process \cite{hohmann2011influence} due to the small width of the PDL layer compared to the commonly-used voxel size of the scans \cite{savignano2019biomechanical,dorow2003finite}. Therefore, for a consistent analysis of different patients and following the literature, the PDL layers of this study are generated by extruding the teeth roots with a uniform thickness of 0.2 mm. Still, further investigation on the PDL is required to study, for example, the effect of nonuniform PDL geometry on the position of the computed CRots since different factors such as the patient’s age and the presence of periodontal disease can influence the shape and thickness of the PDL geometry \cite{li2018orthodontic}, and consequently, the local stress maxima and tooth displacements \cite{hohmann2011influence}. In addition, the same material parameters are used in this study for the teeth, PDLs, and bones of different patients. It should however be noted that the initial and long-term teeth movements can vary from an individual to another based on differences in material properties, the density of the surrounding bone, and the rate of the bone remodeling process, apart from the geometrical differences.

In addition to an accurate geometry reconstruction, utilizing appropriate material models and parameters is important for accurately describing the mechanical behavior of the PDL. Although experimental studies investigating the biomechanical behavior of the PDL in vitro \cite{sanctuary2005vitro} and vivo \cite{keilig2016vivo,yoshida2001vivo,yoshida2001experimental} have indicated an anisotropic nonlinear viscoelastic behavior of the fibrous PDL \cite{qian2009deformation}, computational studies mostly consider simpler material models such as linear elastic, bilinear, and piecewise linear \cite{cattaneo2005finite} models. The linear assumption confines the range of the applied load \cite{kawarizadeh2004correlation}, preventing the simulation of realistic load systems in clinical environments. More complex material models such as hyperplastic and viscoelastic have been used in the literature \cite{qian2009deformation,natali2008visco}, and few studies \cite{ortun2018approach} have considered collagen fibers in the PDL layer to simulate an anisotropic behavior of the PDL layer, allowing it to absorb more energy in compression than in tension and shear \cite{fill2012analytically,komatsu2010mechanical}. However, more investigations are required to see the effectiveness of such complex material models applied to tooth movement modeling using full dentition data in terms of computational complexity and accuracy.

\section{Conclusion}

In this work, three patient-specific computational models of human mandibles, composed of full dentition, PDL layers, and jawbone, were utilized to assess the position of the CRots in different patients’ teeth against varying $M:F$ ratios. The patients’ teeth were aligned for a consistent inter-patient analysis and FE simulations were performed under identical scenarios and boundary conditions for different teeth of each patient to obtain the CRots.

This work benefitted from using more realistic material models and scenarios for better modeling of the forces in clinical environments. A hyperelastic material model was used for the PDL tissue under orthodontic forces which allows for modeling different scenarios without any force constraints. The 3D space was used for representing the positions of the CRots for the exerted force and moment for better analysis and understanding of the results in real-world applications.

The influence of the couple's magnitudes and directions on the positions of the CRots of the patients’ teeth were examined. The results show that the CRot trajectories could follow similar patterns in the corresponding teeth, but any changes in the direction of the force and couple could cause misalignments of the CRot trajectories that could be seen as rotations about the long axis of the tooth.

This work considered a more realistic force system in multi-patient tooth movement modeling by studying the CRot positions. The measured CRot position in relation to the tooth geometry can be used to infer the type of tooth movement, e.g., pure translation, uncontrolled tipping, and crown or root tipping. This, in turn, can be used in treatment planning software to assist clinicians to identify optimal forces required for achieving a desired patient-specific tooth movement.

\section*{Acknowledgments}
The authors would like to thank Dr. Paolo Maria Cattaneo from Melbourne Dental School of the University of Melbourne for his valuable contribution in verifying the segmented CBCT scans used in this study.

\bibliographystyle{unsrt}
\bibliography{references}

\begin{thebibliography}{10}

\bibitem{savignano2019biomechanical}
R~Savignano, R~Valentino, AV~Razionale, A~Michelotti, S~Barone, and
  V~D’Ant{\`o}.
\newblock Biomechanical effects of different auxiliary-aligner designs for the
  extrusion of an upper central incisor: a finite element analysis.
\newblock {\em Journal of Healthcare Engineering}, 2019, 2019.

\bibitem{cervino2020application}
Gabriele Cervino, Luca Fiorillo, Alina~V Arzukanyan, Gianrico Spagnuolo, Paola
  Campagna, and Marco Cicci{\`u}.
\newblock Application of bioengineering devices for stress evaluation in
  dentistry: the last 10 years fem parametric analysis of outcomes and current
  trends.
\newblock {\em Minerva Stomatologica}, 69(1):55--62, 2020.

\bibitem{lavorgna2019reliability}
Luca Lavorgna, Gabriele Cervino, Luca Fiorillo, Giovanni Di~Leo, Giuseppe
  Troiano, Marco Ortensi, Luigi Galantucci, and Marco Cicci{\`u}.
\newblock Reliability of a virtual prosthodontic project realized through a
  {2D} and {3D} photographic acquisition: {A}n experimental study on the
  accuracy of different digital systems.
\newblock {\em International Journal of Environmental Research and Public
  Health}, 16(24):5139, 2019.

\bibitem{sambataro2018upper}
Sergio Sambataro, Gabriele Cervino, Luca Fiorillo, and Marco Cicci{\`u}.
\newblock Upper first premolar positioning evaluation for the stability of the
  dental occlusion: {A}natomical considerations.
\newblock {\em Journal of Craniofacial Surgery}, 29(5):1366--1369, 2018.

\bibitem{gholamalizadeh2020mandibular}
Torkan Gholamalizadeh, Sune Darkner, Paolo~Maria Cattaneo, Peter
  S{\o}ndergaard, and Kenny Erleben.
\newblock Mandibular teeth movement variations in tipping scenario: {A} finite
  element study on several patients.
\newblock {\em arXiv preprint arXiv:2010.05258}, 2020.

\bibitem{nyashin2016centre}
Y~Nyashin, M~Nyashin, M~Osipenko, V~Lokhov, A~Dubinin, F~Rammerstorfer, and
  A~Zhurov.
\newblock Centre of resistance and centre of rotation of a tooth: experimental
  determination, computer simulation and the effect of tissue nonlinearity.
\newblock {\em Computer Methods in Biomechanics and Biomedical Engineering},
  19(3):229--239, 2016.

\bibitem{meyer2010does}
Brandon~N Meyer, Jie Chen, and Thomas~R Katona.
\newblock Does the center of resistance depend on the direction of tooth
  movement?
\newblock {\em American Journal of Orthodontics and Dentofacial Orthopedics},
  137(3):354--361, 2010.

\bibitem{viecilli2013axes}
Rodrigo~F Viecilli, Amanda Budiman, and Charles~J Burstone.
\newblock Axes of resistance for tooth movement: does the center of resistance
  exist in 3-dimensional space?
\newblock {\em American Journal of Orthodontics and Dentofacial Orthopedics},
  143(2):163--172, 2013.

\bibitem{sia2007determining}
Sheau~Soon Sia, Yoshiyuki Koga, and Noriaki Yoshida.
\newblock Determining the center of resistance of maxillary anterior teeth
  subjected to retraction forces in sliding mechanics: an in vivo study.
\newblock {\em The Angle Orthodontist}, 77(6):999--1003, 2007.

\bibitem{sia2009experimental}
SheauSoon Sia, Tatsunori Shibazaki, Yoshiyuki Koga, and Noriaki Yoshida.
\newblock Experimental determination of optimal force system required for
  control of anterior tooth movement in sliding mechanics.
\newblock {\em American Journal of Orthodontics and Dentofacial Orthopedics},
  135(1):36--41, 2009.

\bibitem{yoshida2001experimental}
Noriaki Yoshida, Paul-Georg Jost-Brinkmann, Yoshiyuki Koga, Naofumi Mimaki, and
  Kazuhide Kobayashi.
\newblock Experimental evaluation of initial tooth displacement, center of
  resistance, and center of rotation under the influence of an orthodontic
  force.
\newblock {\em American Journal of Orthodontics and Dentofacial Orthopedics},
  120(2):190--197, 2001.

\bibitem{van2013analytical}
An~Van~Schepdael, Lies Geris, and Jos Vander~Sloten.
\newblock Analytical determination of stress patterns in the periodontal
  ligament during orthodontic tooth movement.
\newblock {\em Medical Engineering \& Physics}, 35(3):403--410, 2013.

\bibitem{provatidis2001analytical}
Christopher~G Provatidis.
\newblock An analytical model for stress analysis of a tooth in translation.
\newblock {\em International Journal of Engineering Science},
  39(12):1361--1381, 2001.

\bibitem{oh20193}
Moon-Bee Oh, Sung-Seo Mo, Chung-Ju Hwang, Chooryung Chung, Ju-Man Kang, and
  Kee-Joon Lee.
\newblock The 3-dimensional zone of the center of resistance of the mandibular
  posterior teeth segment.
\newblock {\em American Journal of Orthodontics and Dentofacial Orthopedics},
  156(3):365--374, 2019.

\bibitem{schneider2002numerical}
J{\"u}rgen Schneider, Martin Geiger, and Franz-G{\"u}nter Sander.
\newblock Numerical experiments on long-time orthodontic tooth movement.
\newblock {\em American Journal of Orthodontics and Dentofacial Orthopedics},
  121(3):257--265, 2002.

\bibitem{schmidt2016comparison}
Falko Schmidt, Martin~Eberhard Geiger, Rudolf J{\"a}ger, and Bernd~Georg
  Lapatki.
\newblock Comparison of methods to determine the centre of resistance of teeth.
\newblock {\em Computer Methods in Biomechanics and Biomedical Engineering},
  19(15):1673--1682, 2016.

\bibitem{melsen2007importance}
Birte Melsen, Paolo~Maria Cattaneo, Michel Dalstra, and David~Christian Kraft.
\newblock The importance of force levels in relation to tooth movement.
\newblock In {\em Seminars in Orthodontics}, pages 220--233. Elsevier, 2007.

\bibitem{liao2016biomechanical}
Zhipeng Liao, Junning Chen, Wei Li, M~Ali Darendeliler, Michael Swain, and Qing
  Li.
\newblock Biomechanical investigation into the role of the periodontal ligament
  in optimising orthodontic force: a finite element case study.
\newblock {\em Archives of Oral Biology}, 66:98--107, 2016.

\bibitem{smith1984mechanics}
Richard~J Smith and Charles~J Burstone.
\newblock Mechanics of tooth movement.
\newblock {\em American Journal of Orthodontics}, 85(4):294--307, 1984.

\bibitem{savignano2016nonlinear}
Roberto Savignano, Rodrigo~F Viecilli, Alessandro Paoli, Armando~Viviano
  Razionale, and Sandro Barone.
\newblock Nonlinear dependency of tooth movement on force system directions.
\newblock {\em American Journal of Orthodontics and Dentofacial Orthopedics},
  149(6):838--846, 2016.

\bibitem{burstone1962biomechanics}
Charles~J Burstone.
\newblock The biomechanics of tooth movement.
\newblock {\em Vistas in Orthodontics}, pages 197--213, 1962.

\bibitem{cattaneo2008moment}
Paolo~M Cattaneo, Michel Dalstra, and Birte Melsen.
\newblock Moment-to-force ratio, center of rotation, and force level: a finite
  element study predicting their interdependency for simulated orthodontic
  loading regimens.
\newblock {\em American Journal of Orthodontics and Dentofacial Orthopedics},
  133(5):681--689, 2008.

\bibitem{savignano2020three}
Roberto Savignano, Rodrigo~F Viecilli, and Udochukwu Oyoyo.
\newblock Three-dimensional nonlinear prediction of tooth movement from the
  force system and root morphology.
\newblock {\em The Angle Orthodontist}, 90(6):811--822, 2020.

\bibitem{fedorov20123d}
Andriy Fedorov, Reinhard Beichel, Jayashree Kalpathy-Cramer, Julien Finet,
  Jean-Christophe Fillion-Robin, Sonia Pujol, Christian Bauer, Dominique
  Jennings, Fiona Fennessy, Milan Sonka, et~al.
\newblock {3D Slicer} as an image computing platform for the quantitative
  imaging network.
\newblock {\em Magnetic Resonance Imaging}, 30(9):1323--1341, 2012.

\bibitem{li2018orthodontic}
Yina Li, Laura~A Jacox, Shannyn~H Little, and Ching-Chang Ko.
\newblock Orthodontic tooth movement: {T}he biology and clinical implications.
\newblock {\em The Kaohsiung Journal of Medical Sciences}, 34(4):207--214,
  2018.

\bibitem{barone2017mechanical}
Sandro Barone, Alessandro Paoli, NERI Paolo, Armando~Viviano Razionale, and
  Michele Giannese.
\newblock Mechanical and geometrical properties assessment of thermoplastic
  materials for biomedical application.
\newblock In {\em Advances on Mechanics, Design Engineering and Manufacturing},
  pages 437--446. Springer, 2017.

\bibitem{seo2021comparative}
Jeong-Hee Seo, Emmanuel Eghan-Acquah, Min-Seok Kim, Jeong-Hyeon Lee, Yong-Hoon
  Jeong, Tae-Gon Jung, Mihee Hong, Won-Hyeon Kim, Bongju Kim, and Sung-Jae Lee.
\newblock Comparative analysis of stress in the periodontal ligament and center
  of rotation in the tooth after orthodontic treatment depending on clear
  aligner thickness—finite element analysis study.
\newblock {\em Materials}, 14(2):324, 2021.

\bibitem{schmidt2010meshmixer}
Ryan Schmidt and Karan Singh.
\newblock Meshmixer: an interface for rapid mesh composition.
\newblock In {\em ACM SIGGRAPH 2010 Talks}, page~6. ACM, 2010.

\bibitem{si2015tetgen}
Hang Si.
\newblock Tetgen, a {D}elaunay-based quality tetrahedral mesh generator.
\newblock {\em ACM Transactions on Mathematical Software (TOMS)}, 41(2):1--36,
  2015.

\bibitem{shewchuk2002good}
Jonathan Shewchuk.
\newblock What is a good linear finite element? interpolation, conditioning,
  anisotropy, and quality measures (preprint).
\newblock {\em University of California at Berkeley}, 73:137, 2002.

\bibitem{liu1994relationship}
Anwei Liu and Barry Joe.
\newblock Relationship between tetrahedron shape measures.
\newblock {\em BIT Numerical Mathematics}, 34(2):268--287, 1994.

\bibitem{misztal2013multiphase}
Marek~Krzysztof Misztal, Kenny Erleben, Adam Bargteil, Jens Fursund,
  Brian~Bunch Christensen, Jakob~Andreas B{\ae}rentzen, and Robert Bridson.
\newblock Multiphase flow of immiscible fluids on unstructured moving meshes.
\newblock {\em IEEE Transactions on Visualization and Computer Graphics},
  20(1):4--16, 2013.

\bibitem{baker1989element}
Timothy~J Baker.
\newblock Element quality in tetrahedral meshes.
\newblock In {\em 7th International Conference on Finite Element Methods in
  Flow Problems}, page 1018, 1989.

\bibitem{freitag1997tetrahedral}
Lori~A Freitag and Carl Ollivier-Gooch.
\newblock Tetrahedral mesh improvement using swapping and smoothing.
\newblock {\em International Journal for Numerical Methods in Engineering},
  40(21):3979--4002, 1997.

\bibitem{caendish1985apporach}
James~C Caendish, David~A Field, and William~H Frey.
\newblock An approach to automatic three-dimensional finite element mesh
  generation.
\newblock {\em International Journal for Numerical Methods in Engineering},
  21(2):329--347, 1985.

\bibitem{maas2012febio}
Steve~A Maas, Benjamin~J Ellis, Gerard~A Ateshian, and Jeffrey~A Weiss.
\newblock {FEBio}: finite elements for biomechanics.
\newblock {\em Journal of Biomechanical Engineering}, 134(1), 2012.

\bibitem{zmudzki2008stress}
J~{\.Z}mudzki and W~Chladek.
\newblock Stress present in bone surrounding dental implants in {FEM} model
  experiments.
\newblock {\em Journal of Achievements in Materials and Manufacturing
  Engineering}, 27(1):71--74, 2008.

\bibitem{roda2021computerized}
Victor Roda-Casanova, {\'A}lvaro Zubizarreta-Macho, Francisco Sanchez-Marin,
  {\'O}scar Alonso~Ezpeleta, Alberto Albaladejo~Mart{\'\i}nez, and Agust{\'\i}n
  Galparsoro~Catal{\'a}n.
\newblock Computerized generation and finite element stress analysis of
  endodontic rotary files.
\newblock {\em Applied Sciences}, 11(10):4329, 2021.

\bibitem{ortun2018approach}
J~Ort{\'u}n-Terrazas, J~Cego{\~n}ino, U~Santana-Pen{\'\i}n, U~Santana-Mora, and
  A~P{\'e}rez del Palomar.
\newblock Approach towards the porous fibrous structure of the periodontal
  ligament using micro-computerized tomography and finite element analysis.
\newblock {\em Journal of the Mechanical Behavior of Biomedical Materials},
  79:135--149, 2018.

\bibitem{FEBioUserManual}
SA~Maas, D~Rawlins, JA~Weiss, and GA~Ateshian.
\newblock {FEBio} 2.8 user manual, 2018.

\bibitem{ziegler2005numerical}
A~Ziegler, L~Keilig, A~Kawarizadeh, A~J{\"a}ger, and C~Bourauel.
\newblock Numerical simulation of the biomechanical behaviour of multi-rooted
  teeth.
\newblock {\em The European Journal of Orthodontics}, 27(4):333--339, 2005.

\bibitem{qian2009deformation}
Lihe Qian, Mitsugu Todo, Yasuyuki Morita, Yasuyuki Matsushita, and Kiyoshi
  Koyano.
\newblock Deformation analysis of the periodontium considering the
  viscoelasticity of the periodontal ligament.
\newblock {\em Dental Materials}, 25(10):1285--1292, 2009.

\bibitem{cattaneo2005finite}
PM~Cattaneo, M~Dalstra, and B~Melsen.
\newblock The finite element method: {A} tool to study orthodontic tooth
  movement.
\newblock {\em Journal of Dental Research}, 84(5):428--433, 2005.

\bibitem{uhlir2016biomechanical}
Richard Uhlir, Virginia Mayo, Pei~Hua Lin, Si~Chen, Yan-Ting Lee, Garland
  Hershey, Feng-Chang Lin, and Ching-Chang Ko.
\newblock Biomechanical characterization of the periodontal ligament:
  {O}rthodontic tooth movement.
\newblock {\em The Angle Orthodontist}, 87(2):183--192, 2016.

\bibitem{chen1992object}
Yang Chen and G{\'e}rard Medioni.
\newblock Object modelling by registration of multiple range images.
\newblock {\em Image and Vision Computing}, 10(3):145--155, 1992.

\bibitem{besl1992method}
Paul~J Besl and Neil~D McKay.
\newblock A method for registration of {3-D} shapes.
\newblock {\em IEEE Transactions on Pattern Analysis and Machine Intelligence},
  14(2):239--256, 1992.

\bibitem{myronenko2010point}
Andriy Myronenko and Xubo Song.
\newblock Point set registration: {C}oherent point drift.
\newblock {\em IEEE Transactions on Pattern Analysis and Machine Intelligence},
  32(12):2262--2275, 2010.

\bibitem{kawarizadeh2004correlation}
Afshar Kawarizadeh, Christoph Bourauel, Dongliang Zhang, Werner G{\"o}tz, and
  Andreas J{\"a}ger.
\newblock Correlation of stress and strain profiles and the distribution of
  osteoclastic cells induced by orthodontic loading in rat.
\newblock {\em European Journal of Oral Sciences}, 112(2):140--147, 2004.

\bibitem{hohmann2011influence}
Ansgar Hohmann, Cornelia Kober, Philippe Young, Christina Dorow, Martin Geiger,
  Andrew Boryor, Franz~Martin Sander, Christian Sander, and Franz~G{\"u}nter
  Sander.
\newblock Influence of different modeling strategies for the periodontal
  ligament on finite element simulation results.
\newblock {\em American Journal of Orthodontics and Dentofacial Orthopedics},
  139(6):775--783, 2011.

\bibitem{dorow2003finite}
Christina Dorow, Juergen Schneider, and Franz~G Sander.
\newblock Finite element simulation of in vivo tooth mobility in comparison
  with experimental results.
\newblock {\em Journal of Mechanics in Medicine and Biology}, 3(01):79--94,
  2003.

\bibitem{sanctuary2005vitro}
Colin~S Sanctuary, HW~Anselm Wiskott, Jorn Justiz, John Botsis, and Urs~C
  Belser.
\newblock In vitro time-dependent response of periodontal ligament to
  mechanical loading.
\newblock {\em Journal of Applied Physiology}, 99(6):2369--2378, 2005.

\bibitem{keilig2016vivo}
L~Keilig, M~Drolshagen, KL~Tran, I~Hasan, S~Reimann, J~Deschner, KT~Brinkmann,
  R~Krause, M~Favino, and C~Bourauel.
\newblock In vivo measurements and numerical analysis of the biomechanical
  characteristics of the human periodontal ligament.
\newblock {\em Annals of Anatomy-Anatomischer Anzeiger}, 206:80--88, 2016.

\bibitem{yoshida2001vivo}
Noriaki Yoshida, Yoshiyuki Koga, Chien-Lun Peng, Eiji Tanaka, and Kazuhide
  Kobayashi.
\newblock In vivo measurement of the elastic modulus of the human periodontal
  ligament.
\newblock {\em Medical Engineering \& Physics}, 23(8):567--572, 2001.

\bibitem{natali2008visco}
Arturo~N Natali, Emanuele~L Carniel, Piero~G Pavan, Franz~G Sander, Christina
  Dorow, and Martin Geiger.
\newblock A visco-hyperelastic-damage constitutive model for the analysis of
  the biomechanical response of the periodontal ligament.
\newblock {\em Journal of Biomechanical Engineering}, 130(3), 2008.

\bibitem{fill2012analytically}
Ted~S Fill, Roger~W Toogood, Paul~W Major, and Jason~P Carey.
\newblock Analytically determined mechanical properties of, and models for the
  periodontal ligament: critical review of literature.
\newblock {\em Journal of Biomechanics}, 45(1):9--16, 2012.

\bibitem{komatsu2010mechanical}
Koichiro Komatsu.
\newblock Mechanical strength and viscoelastic response of the periodontal
  ligament in relation to structure.
\newblock {\em Journal of Dental Biomechanics}, 2010, 2010.

\end{thebibliography}

\end{document}